\documentclass[a4paper,11pt]{article}
\pdfoutput=1

\usepackage{jcappub} 
\usepackage[T1]{fontenc}
\usepackage{ulem}

\newcommand{\citeR}[1]{Ref.~\cite{#1}}

\title{\boldmath Information Gains from Cosmological Probes}

\author[a,b,c]{S. Grandis,}
\author[a]{S. Seehars,}
\author[a]{A. Refregier,}
\author[a]{A. Amara,}
\author[a]{A. Nicola}
\affiliation[a]{Department of Physics, ETH Z\"urich, Wolfgang-Pauli-Strasse 27, CH-8093 Zurich, Switzerland}
\affiliation[b]{Faculty of Physics, Ludwig-Maximilian University, Scheinerstr. 1, 81679 Munich, Germany}
\affiliation[c]{Excellence Cluster Universe, Boltzmannstr. 2, 85748 Garching, Germany}

\emailAdd{grandiss@phys.ethz.ch}
\emailAdd{grandis@usm.lmu.de}
\emailAdd{sebastian.seehars@phys.ethz.ch}
\emailAdd{alexandre.refregier@phys.ethz.ch}
\emailAdd{adam.amara@phys.ethz.ch}
\emailAdd{andrina.nicola@phys.ethz.ch}

\abstract{
In light of the growing number of cosmological observations, it is important to develop versatile tools to quantify the constraining power and consistency of cosmological probes. Originally motivated from information theory, we use the relative entropy to compute the information gained by Bayesian updates in units of bits. This measure quantifies both the improvement in precision and the 'surprise', i.e. the tension arising from shifts in central values. Our starting point is a WMAP9 prior which we update with observations of the distance ladder, supernovae (SNe), baryon acoustic oscillations (BAO), and weak lensing as well as the 2015 Planck release. We consider the parameters of the flat $\Lambda$CDM concordance model and some of its extensions which include curvature and Dark Energy equation of state parameter $w$. We find that, relative to WMAP9 and within these model spaces, the probes that have provided the greatest gains are Planck (10 bits), followed by BAO surveys (5.1 bits) and SNe experiments (3.1 bits). The other cosmological probes, including weak lensing (1.7 bits) and {$\rm H_0$} measures (1.7 bits), have contributed information but at a lower level. Furthermore, we do not find any significant surprise when updating the constraints of WMAP9 with any of the other experiments, meaning that they are consistent with WMAP9. However, when we choose Planck15 as the prior, we find that, accounting for the full multi-dimensionality of the parameter space, the weak lensing measurements of CFHTLenS produce a large surprise of 4.4 bits which is statistically significant at the 8 $\sigma$ level. We discuss how the relative entropy provides a versatile and robust framework to compare cosmological probes in the context of current and future surveys.
}

\notoc
\begin{document}
\maketitle
\flushbottom

\section{Introduction}

Cosmology has been the subject of substantial observational progress in the last decade. A large number of probes are now combined to constrain the content and evolution of the Universe. This has led to the emergence of the $\Lambda$CDM concordance model which has been found to be mostly consistent with all the observations. Due to the complexity and diversity of the probes, it is important to be able to quantify and compare their power and consistency in a general and robust framework.

In an earlier work by~\citeR{seehars1}, we investigated such a framework based on the relative entropy, or Kullback-Leibler divergence~\cite{kullbackleibler}, motivated from information theory (see also related work by~\citeR{paykari, modelbreaking, march, kunz}). This divergence measure quantifies the information gain when incorporating data from a given experiment by performing a Bayesian update from prior to posterior constraints. It is sensitive to both the reduction of statistical errors and shifts in the central values of the model parameters. In~\citeR{seehars1, seehars2}, we focused on a historical sequence of cosmic microwave background (CMB) experiments and their constraints on flat $\Lambda$CDM. The information gain captured the improvement in precision as CMB data got better, but also detected tensions between the individual data releases of the WMAP team or the WMAP and Planck 2013 constraints, for example.

Building on the work on CMB constraints for flat $\Lambda$CDM cosmologies~\cite{seehars1}, we apply the relative entropy to a larger number of cosmological probes and to a wider range of cosmological models. To do so, we update the CMB constraints~\cite{wmap1, wmap2, planck13, planck15, planck13_cp, planck_likelis} with constraints on the Hubble constant ($H_0$) from distance ladder measurements~\cite{efstathiou, riess, riess14}, Supernovae Type Ia (SNe) constraints~\cite{conley, suzuki, betoule}, baryon acoustic oscillations (BAO) measurements~\cite{beutler, anderson, kazin, ross, aubourg, delubac}, weak lensing constraints derived from measurements of the cosmic shear correlation function~\cite{kilbinger, mandelbaum, heymans_tomographic, Kitching, fu, 2015arXiv150705552T, maccrann}, and CMB lensing measurements~\cite{planck_lensing}. In addition to a flat $\Lambda$CDM model, we also consider non-flat $\Lambda$CDM and flat $w$CDM\footnote{Flat $w$CDM is a model with a Dark Energy component that has a constant equation of state parameter $w$ which is allowed to deviate from that of a cosmological constant ($w = -1$), but is constant in time.} cosmologies. In cosmological models beyond flat $\Lambda$CDM, however, degeneracies between parameters lead to constraints that are not well approximated by Gaussian distributions. To estimate the relative entropy from samples of such generic distributions, we develop and implement a shapelet based method.

From these analyses, we study the information gains of CMB experiments and the other cosmological probes. We first consider $\Lambda$CDM as the fiducial model and then investigate how the resulting ranking of the individual probes depends on the model extensions. We also quantify possible tensions between the CMB and other probes. This provides an overview of the current state of observational cosmology through the relative entropy framework.

This work is organised as follows. In Section~\ref{sec:method} we describe the methods used to compute the information gain. Section~\ref{sec:probes} presents the different cosmological probes, motivates the choices of data sets used as representatives of each probe and discusses the implementations of their likelihoods. Thereafter, we present and discuss the results of our analysis in Sections~\ref{sec:results} and~\ref{sec:conc}.

\newpage

\section{The Relative Entropy}\label{sec:method}
Throughout this paper, constraints from cosmological probes are presented as posterior probability distribution functions (PDF) on the space of cosmological parameters. In this section we outline the concepts we use to analyse such PDFs.

\subsection{Theory}\label{sec:theor_back}
Given a prior $q(\pmb{\theta})$ on a set of parameters $\pmb{\theta}$, and a likelihood $l(\pmb{\theta}) = \mathcal{L}(\text{data}|\pmb{\theta})$, we can update the prior to the posterior $p(\pmb{\theta})$ using Bayes' Theorem by computing 

\begin{equation}\label{eq:bayes}
p(\pmb{\theta}) = \frac{l(\pmb{\theta})}{\langle l(\pmb{\theta}) \rangle_q} \, q(\pmb{\theta}),
\end{equation}
where $\langle l(\pmb{\theta}) \rangle_q = \int d^d\pmb{\theta}\, q(\pmb{\theta}) l(\pmb{\theta})$ is the evidence, ensuring the normalisation of the posterior. 

Differences between the two distributions $p(\pmb{\theta})$ and $q(\pmb{\theta})$ can be measured by the Kullback-Leibler divergence~\cite{kullbackleibler}, hereafter called \emph{relative entropy}
\begin{equation}\label{eq:def_rel_ent}
D(p||q) = \int d^d \pmb{\theta} \, p(\pmb{\theta}) \, \ln \Big( \frac{p(\pmb{\theta})}{q(\pmb{\theta})} \Big).
\end{equation}
The relative entropy is positive and invariant under remapping of parameters and marginalisation of unconstrained parameters (see details in Appendix~\ref{sec:entr_props}). It quantifies the information gain from one distribution to another (for more details see e.g.~\citeR{info_theory}). As information is usually measured in bits, we need to normalise the relative entropy by a factor of $\ln 2$. To create an intuition of how much 1 bit is, consider a one-dimensional gaussian: 1 bit of information can be achieved by shifting its central value by 1.18 $\sigma$, or by reducing its standard deviation by $68 \%$, as can be computed from Eq. \ref{eq:gauss_entr}. 

Thus, we define the \emph{information gain} of an update $q \rightarrow p$ as $D(p||q)/\ln 2$, measured in bits. Naturally, the information gain inherits the useful properties of the relative entropy shown in Appendix~\ref{sec:entr_props}, which qualifies it as a robust way of assessing the updating power of a cosmological probe.

\subsection{Estimator}\label{sec:gener_entr}
To estimate the relative entropy in practice, we first note that~\eqref{eq:def_rel_ent} can be expressed as the difference between two expectation values
\begin{equation}\label{eq:estimate}
\mathcal{D}(p||q) = \bigg \langle \ln \left( \frac{p}{q} \right) \bigg \rangle_p = \big \langle \ln p \big \rangle_p - \big \langle \ln q \big \rangle_p.
\end{equation}
Given samples from $p$ and $q$, these two expectation values can in principle be computed with a Monte Carlo (MC) integral. The numerical value that is typically available at each sample point, however, is the numerical value of the likelihood up to a normalisation constant rather than the numerical value of posterior and prior. Estimation of this normalisation constant and marginalisation of the likelihood value over nuisance parameters is computationally expensive. To evaluate equation~\eqref{eq:estimate}, we hence not only need to estimate the average over $p$, but we must also approximate the distributions $p$ and $q$ with reconstructions $\hat p$ and $\hat q$ based on the sample. Let $\pmb{\theta}_i^p$ (for $i = 1,\dots,N$) be $N$ samples drawn from $p$. An estimator for the relative entropy can be constructed as
\begin{equation}
\hat D = \frac{1}{N} \sum_{i=1}^N \left[ \ln \hat p(\pmb{\theta}_i^{p}) - \ln \hat q(\pmb{\theta}_i^{p}) \right].
\end{equation}

To estimate the reconstructed distributions $\hat p$ and $\hat q$, nearest neighbour approaches have been proposed by~\citeR{nearest1, nearest2}. However, estimating $\ln \hat q(\pmb{\theta}_i^{p})$ is difficult, as any finite error in $\hat q = q + \delta q$ propagates to $\delta \ln \hat q(\pmb{\theta}_i^{p})= \delta q / q(\pmb{\theta}_i^{p})$. If the overlap between $p$ and $q$ is small, then $q(\pmb{\theta}_i^{p})$ is small, increasing the noise in the estimator $\hat D$, as noted by~\citeR{nearest1}.

\begin{figure}[t]
    \centering
    \includegraphics[width=\textwidth]{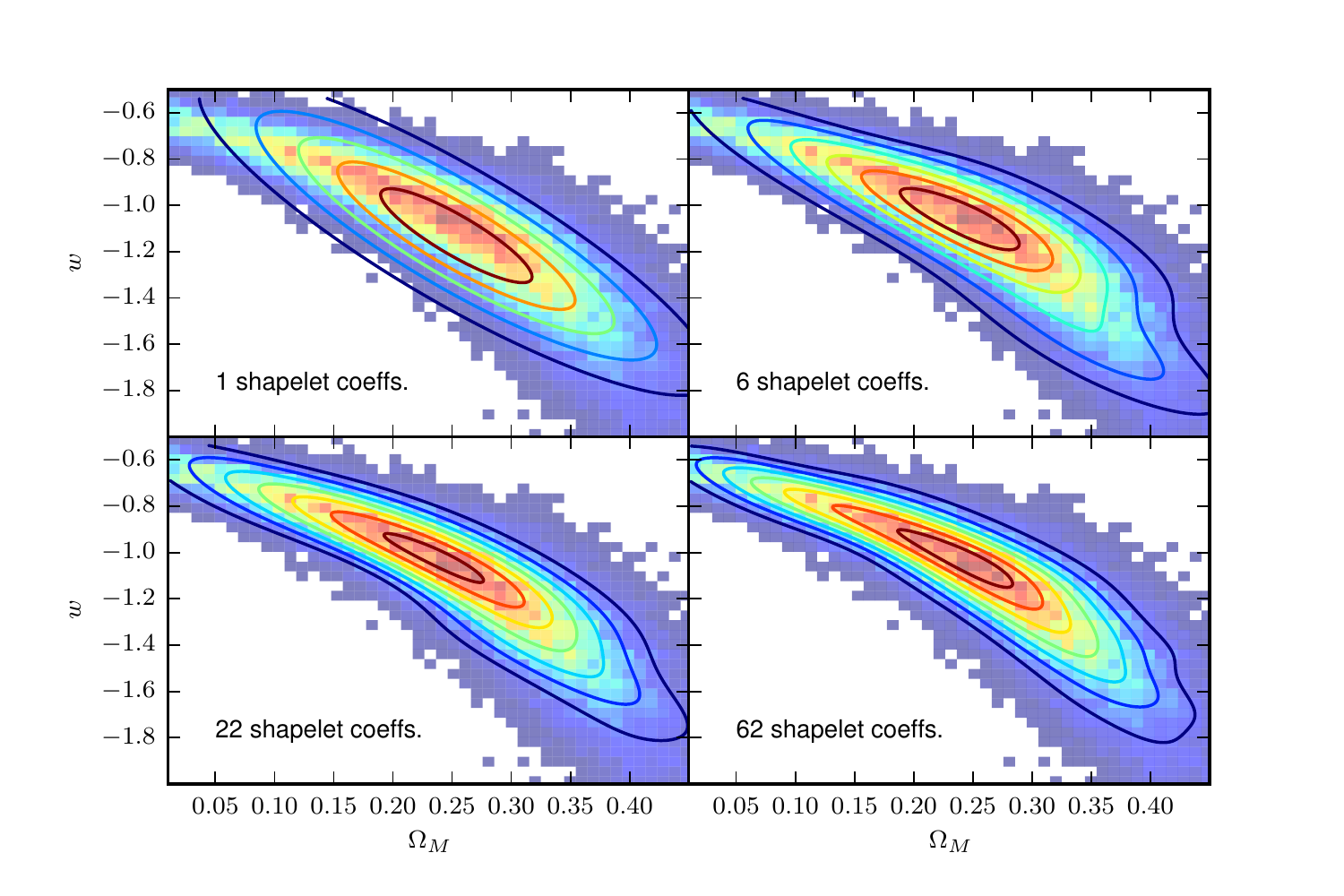}
    \caption{\label{fig:shapelets} Shapelet reconstruction of the non-Gaussian $\Omega_M$, $w$ constraints from SNe data for different numbers of shapelet coefficients. Estimates for the shapelet coefficients, and errors on these estimates, are calculated from the sample. Thus, applying a signal to noise cut $\nu$ allows us to control the number of shapelet coefficients (cf. Appendix~\ref{sec:shapelets}). If only the coefficient with the largest signal to noise is used (top left), the shapelet reconstruction is a Gaussian. Adding further shapelet coefficients improves the approximation.  We show contours of the reconstruction on top of a histogram of the samples used to estimate the shapelet coefficients for different signal to noise cuts (top left: $\nu=200$, top right: $\nu=60$, bottom left: $\nu=25$, bottom right: $\nu=11$).}
\end{figure}

We alleviate this effect by constructing perturbations around a Gaussian with the shapelet method described in Appendix~\ref{sec:shapelets}. Inspired from the shapelet reconstructions of galaxy images presented by~\citeR{shapelets}, we expand our PDF around a Gaussian in a so called Gram-Charlier Series~\cite{gram, charlier1, charlier2, hermites} or Edgeworth Expansion~\cite{edgeworth}. In~\citeR{edgeworth1, edgeworth2, edgeworth3} only the first terms of such an expansion are considered for the analytical computation of the relative entropy between one-dimensional distributions. Generalising this method to higher dimensions is however non-trivial. We hence preferred to compute all the coefficients and to reconstruct the PDFs by means of Gram-Charlier Series, hereafter called shapelet reconstruction. As an example, Figure~\ref{fig:shapelets} shows the reconstruction of the $\Omega_M$, $w$ degeneracy of the SNe constraints, obtained when using different number of shapelet coefficients. Besides regularising the tails this method has the advantage that, as for galaxy images, it compresses the distribution into a small set of shapelet coefficients, speeding up the evaluation of the reconstruction. 

\subsection{Gaussian Approximation}\label{sec:gauss_entr}
If $p$ and $q$ are multivariate normal distributions with means $\pmb{\mu}_p$ and $\pmb{\mu}_q$, and covariances $C_p$ and $C_q$ respectively, the relative entropy is analytic and given by the following expression (see e.g. \citeR{seehars1})
\begin{equation} \label{eq:gauss_entr}
        D(p||q) = \frac{1}{2} (\pmb{\mu}_p-\pmb{\mu}_q)^T C_q^{-1} (\pmb{\mu}_p-\pmb{\mu}_q) -\frac{1}{2} \ln \Bigg( \frac{\det C_p}{\det C_q} \Bigg )  + \frac{1}{2} \text{tr} \Big \{ C_q^{-1}(C_p -C_q) \Big \}.
\end{equation}
The second term expresses the information gain due to a tightening of the contours, as it compares the covariance determinants of prior and posterior. The second term is hence related to the Dark Energy Task Force~\cite{DEFT} figure of merit, which is defined as the inverse covariance determinant in the Dark Energy parameters. The third term is related to an alternative figure of merit, proposed by~\citeR{bayes_in_cosmo, paykari}, which uses the traces of the covariance matrices rather than their determinants. In the combination given above, the sum of these terms expresses the information gain due to the change of covariance matrices, as shown in~\citeR{modelbreaking, paykari, seehars1}. 

The first term is driven by a shift between prior and posterior means and is therefore related to tensions in the update. For the case of a Gaussian prior and likelihood and a linear model, it is shown in~\citeR{seehars1} that this term follows a generalised $\chi^2$ distribution and is expected to have a non-zero expectation value. \citeR{seehars1} hence defines the \emph{surprise} $S$ by subtracting the expected value from the above term
\begin{equation}\label{eq:S}
   S = \frac{1}{2}\pmb{\Delta \mu}^T C_q^{-1} \pmb{\Delta \mu} - \frac{1}{2}\big\langle\pmb{\Delta \mu}^T C_q^{-1} \pmb{\Delta \mu}\big \rangle.
\end{equation}
If the surprise is positive, $S > 0$, the means shifted more than expected. Consequently the update introduced tension which the surprise quantifies in bits. A negative surprise $S < 0$, on the other hand, indicates that the means shifted less than expected. In a similar fashion, \citeR{seehars1, seehars2} also derive the expected variation $\sigma(D)$ of the relative entropy $D$ around its expected value. In the following, we speak of a significant surprise if $S$ is large compared to $\sigma(D)$. Significant surprises could be due to residual systematic effects in the data, or could indicate tensions originating from new physics. For a detailed discussion of the surprise $S$, we refer to~\citeR{seehars1}, as we adopted their definitions. 

As discussed by \citeR{seehars2}, the major advantage of the surprise is twofold: first, it takes account of the full multidimensional overlap between the distribution and is thus not affected by projection effects which can reduce the degree of tension inferred from an analysis of the marginalised constraints. Second, it allows us to calculate the level of significance of a tension, as it has an expected value of zero and a known scatter. This is usually not possible with other measures of data set agreement, such as evidence ratios or the analysis of marginalised constraints. Although for Gaussian constraints the distance in means in units standard deviations is a valuable measure of agreement, the surprise calculations allow us to estimate the how significant such a shift is. Thus, one should be careful when comparing the significances of surprises to differences of means in units of standard deviations, as these to quantities are both reported in units of `$\sigma$'s, but are different measures.

It worth noting, that the relative entropy depends on the specific data set used to update from the prior to the posterior. For this reason, in the following we present the data sets we choose as representatives of the present state of cosmological probes. Our choices follow mostly the Planck Collaboration \cite{planck13, planck15}.

\section{Cosmological Probes}\label{sec:probes}

In this work, we generate samples of prior and posterior using the parallelised Monte Carlo Markov Chain (MCMC) sampler \texttt{CosmoHammer}~\cite{cosmohammer}. In the following we discuss the observational details of each probe as well as the implementation of the necessary modules,  as the numerical results presented in the course of this paper depend on the specific choices of data sets we used.

\subsection{Cosmic Microwave Background}\label{sec:CMB_probes}

\paragraph{WMAP}
As a prior for most of this work, we will use the constraints coming from the 9-year Wilkinson Microwave Anisotropy Probe (WMAP9)~\cite{wmap1, wmap2}. We use both the temperature and polarisation anisotropy measurements. The implementation is done using the \texttt{CosmoHammer} plugins\footnote{https://github.com/cosmo-ethz/CosmoHammerPlugins} for \texttt{PyCamb}, a python wrapper of \texttt{Camb}~\cite{Lewis:2008wx}, and for the official WMAP 9 likelihood.

\paragraph{Planck} Besides WMAP9, we also analyse the constraints coming from the temperature and polarisation anisotropy measurements of the Planck Collaboration~\cite{planck15}, hereafter called Planck15. We base our analysis on the published chains\footnote{http://pla.esac.esa.int/pla/\#results: we used the \texttt{plikHM\_TTTEEE\_lowTEB} chains.}. When we use the Planck data to update the WMAP 9 constraint, we apply the scheme presented by \citeR{seehars1}. When used as a prior, we take the Planck15 constraints to be a multivariate normal distribution with mean and covariance estimated from the chains. Note that although the Planck Collaboration \cite[pag. 15]{planck15} does not recommend the use of the TE and EE spectra, as they might be affected by a temperature to polarisation leakage, considering the very good agreement \cite[Fig. 6, Tab. 5, Fig. 20]{planck15} between the TT and TT,TE,EE constraints, we take the Planck measurement at face value and base our analysis on both the temperature and the polarisation measurements. In Appendix \ref{sec:planck_pol}, we discuss the impact of the polarisation data on our conclusions. 

\subsection{Geometrical Probes}\label{sec:geom_probes}
It is well known (see e.g.~\citeR{howlett}) that when extending the flat $\Lambda$CDM model to non-flat geometries or arbitrary, but constant equation of state parameter of Dark Energy, the CMB constraints display degeneracies. These degeneracies can be broken by geometrical probes, such as H$_0$ measurements, Supernovae Type Ia observations, or Baryonic Acoustic Oscillations data (as shown by e.g.~\citeR{efstathiou2}). We implement these probes as follows. 

\paragraph{Hubble Constant}
There is a long and sometimes controversial history of Hubble constant measurements (e.g.~\citeR{h0history0, h0history00, h0history01, hubble, h0history1,h0history2}) and ongoing efforts to improve these measurements (e.g.~\citeR{H0concordance, riess, aubourg, efstathiou}). Following the choice of Planck15~\cite{planck15} we used the recent measurement based on the NGC 4258 maser distance conducted by~\citeR{efstathiou}. We assume that the H$_0$-likelihood is a Gaussian with mean and standard deviation given by $H_0 = 70.6 \text{ km/s/Mpc}$ and $\sigma = 3.3 \text{ km/s/Mpc}$, respectively. Given that different values for $\rm H_0$ can be found in the literature (e.g. \citeR{riess, riess14}), we show in Appendix \ref{sec:h0} how the specific choice of dataset affects the estimated information gains.

%\footnote{\red{It is well known that different values for $\rm H_0$ can be found in the literature. For instance \citeR{H0concordance} uses  $H_0 = 73.0 \text{ km/s/Mpc}$ and $\sigma = 2.4 \text{ km/s/Mpc}$, reported by \citeR{riess14}. To check that our main conclusions do depend on these specific choices, we performed our analysis also with the \citeR{riess14} value and present the results in Appendix \ref{sec:h0}. The main plots, however, follow the choices made by the Planck Collaboration.}}

\paragraph{Supernovae Type Ia} 
The present work is based on the SNLS+SDSS+HST SNe compilation, analysed by~\citeR{conley}. Using the publicly available data and covariance matrix, we implement the likelihood discussed in~\citeR{conley} as a \texttt{CosmoHammer} module in \texttt{python}. This requires the computation of the "Hubble free" luminosity distance with \texttt{PyCamb}. We took explicit account of systematics by marginalising analytically over the absolute magnitude $\mathcal{M}$ and numerically over the nuisance parameters $\alpha$ and $\beta$ accounting for the stretch-luminosity and colour-luminosity relationship. As for $\rm H_0$, different SNe data sets exist in the literature (e.g. the Union 2.1 sample by \citeR{suzuki}, or the reanalysis and recalibration of the sample of \citeR{conley} by \citeR{betoule}). Following the choice of \citeR{planck13}, we used the calibration by \citeR{conley}.

\paragraph{Baryonic Acoustic Oscillations} It is customary to report isotropic BAO results in terms of $D_V(z)\, r_\text{fid} / r_\text{drag}$. The fiducial sound horizon at recombination $r_\text{fid}$ is fixed by the fiducial cosmology that was assumed during the analysis of the data. The cosmology dependent sound horizon at recombination is denoted by $r_\text{drag}$. Finally, $D_V(z)$ is given by
\begin{equation}
D_V(z) = \Big \{  (1+z)^2 D_A^2(z) \frac{cz}{H(z)} \Big \}^{\frac{1}{3}},
\end{equation}
where $D_A(z)$ is the angular diameter distance. We compute $r_\text{drag}$ and $D_A(z)$ with \texttt{Camb}.

The data compilation we use is presented in Table~\ref{tab:baos} and is taken from \citeR{planck15}. Given the data, the theory prediction and the covariance, we assume a Gaussian likelihood. For simplicity, we ignore the possible correlations arising from the overlap between the SDSS III field and the WiggleZ field. Following \citeR{planck15, aubourg, dark_current}, we choose this BAO data set as representative of the current constraints. However, as \citeR{planck15}, we did not consider Ly$\alpha$ forest BAO measurements, such as \citeR{delubac}.

\begin{table}[t]
\centering
\caption{BAO measurements used in the present work. The measurements from 6dFGS and BOSS are taken in the form used by~\citeR{H0concordance}. We ignore the correlations from the overlap between the SDSS field and the WiggleZ field.}
\label{tab:baos}
\vspace{10pt}
\begin{tabular}{|cccccc|}
\hline
Survey & $D_V(z)\, r_\text{fid} / r_\text{drag}$ & $z$ & $r_\text{fid}$ in Mpc & Error & Reference \\
\hline
6dFGS & 456 Mpc & 0.106 & 153.19 & $\pm 20$ Mpc & \citeR{beutler} \\ 

BOSS in SDSS III & 1264 Mpc & 0.32 & 153.19 & $\pm 25$ Mpc & \citeR{anderson} \\
 & 2056 Mpc & 0.57 & & $\pm 20$ Mpc & \\

WiggleZ & 457 Mpc & 0.106 & 148.6 & covariance & \citeR{kazin} \\
 & 1716.4 Mpc & 0.44 & & matrix &  \\
 & 2220.8 Mpc & 0.6 & & given &  \\

SDSS DR7 & 664 Mpc & 0.15 & 148.69 & $\pm 25$ Mpc & \citeR{ross} \\
\hline
\end{tabular}
\end{table}
 
\paragraph{Relative BAOs} 
BAOs can be used as a standard ruler whose size is well constrained by the CMB (cf. \citeR{aubourg}). When computing the information gain from BAOs it is thus natural to ask how much information comes from the CMB calibration of $r_\text{drag}$, and how much information is yielded by the distance measurement itself, as proposed by~\citeR{rel_BAO}. To disentangle the effect of the CMB calibration, we define relative BAOs by allowing $r_\text{drag}$ to vary independently of the other cosmological parameters. In this implementation, BAOs are thus a relative distance measure. In other words, relative BAOs are standard rulers of unknown intrinsic size. This allows us to asses the constraining power of BAOs in more detail.
 
\subsection{Large-scale Structure Probes}

Besides geometrical probes, we also consider large-scale structure (LSS) probes. Note that BAOs are derived from LSS observations. However, we prefer to interpret them as distance measures. In the following, we restrict our LSS analysis to CMB lensing and weak lensing with details given below.

\paragraph{CMB Lensing}
For CMB lensing measurements, we use the publicly available chains presented in~\citeR{planck15}, which give the joint constraints from the Planck15 temperature and polarisation anisotropy measurements and the CMB lensing measurements, discussed in~\citeR{planck_lensing}. We refer to this probe as \emph{Planck15 + CMB lens}.

\paragraph{Weak Gravitational Lensing}
As a further, independent cosmological probe we use the constraints coming from the weak lensing shear correlation function measured by the Canada-France-Hawaii Telescope Lensing Survey (CFHTLenS)~\cite{Heymans}. Following the cosmological analysis by~\citeR{kilbinger}, we use the 2D cosmic shear correlation functions $\xi_{\pm}$ for which the data and the covariance matrix are publicly available\footnote{http://www.cfhtlens.org/astronomers/cosmological-data-products}. For simplicity, we ignore the cosmological parameter dependence of the covariance matrix. As in the original analysis by \citeR{kilbinger}, the tomographic analysis by \citeR{heymans_tomographic}, the 3d analysis by \citeR{Kitching}, and the higher moments analysis by \citeR{fu}, we assume a Gaussian likelihood. We apply no cuts at small angles, thus considering the full shear correlation of CFHTLenS. As \citeR{kilbinger} in the original 2D shear analysis, we do not sample any nuisance parameter parametrising possible systematics. For a detailed discussion of the impact of the systematics on the CFHTLenS constraints and its consistency with other probes, see \citeR{maccrann, WL_bmodes, WL_systematics, WL_systematics_2, WL_systematics_3}.

The theory prediction of the correlation function is implemented with \texttt{PyCosmo}~\cite{pycosmo} as follows. The matter power spectrum is computed using the Eisenstein \& Hu transfer function~\cite{eisensteinhu} and the non-linear corrections are implemented using the revised \texttt{HALOFIT}~\cite{halofit,takahashi}. Given the power spectrum and the redshift distribution of CFHTLenS, we compute the cosmic shear correlation function with the Limber approximation~\cite{limber}.

\section{Results}\label{sec:results}

Using the likelihoods discussed in Section~\ref{sec:probes}, we compute the posteriors for four combinations of priors and models. Table~\ref{tab:likelis} describes the models, the priors, the sampled parameters, and the likelihood we use for our updates. Detailed Tables of the information gains for each individual update are shown in Appendix~\ref{sec:full}. 

\begin{table}[t]
\centering

\caption{Models, priors, parameters and likelihood combinations used for this work.}
\label{tab:likelis}
\vspace{10pt}
\begin{tabular}{|cccc|}
\hline
Model & Prior & Parameters & Updates \\
\hline
flat $\Lambda$CDM & WMAP9 & $(H_0, \Omega_{b}\,h^2, \Omega_{dm}\,h^2, A_s, n_s)$ & SNe, H0, BAOs, rel. BAOs,\\
 & & & CHFTLenS, Planck15\dag\\

non-flat $\Lambda$CDM & WMAP9 & $(H_0, \Omega_{b}\,h^2, \Omega_{dm}\,h^2, A_s, n_s, \Omega_K)$ & SNe, H0, BAOs, rel. BAOs,\\
 & & & CHFTLenS, Planck15\dag, \\

flat $w$CDM & WMAP9 & $(H_0, \Omega_{b}\,h^2, \Omega_{dm}\,h^2, A_s, n_s, w_0)$ & SNe, H0, BAOs, rel. BAOs,\\
 & & & CHFTLenS, Planck15\dag \\

flat $\Lambda$CDM & Planck15 & $(H_0, \Omega_{b}\,h^2, \Omega_{dm}\,h^2, A_s, n_s)$ & SNe, H0, BAOs\dag, rel. BAOs,\\
 & & & CHFTLenS, CMB lens.\dag \\\hline
\multicolumn{4}{l}{{\footnotesize\dag We rely on publicly available chains for these updates.}}
\end{tabular}
\end{table}

As a fiducial configuration we adopt WMAP9 priors within a flat $\Lambda$CDM cosmology by sampling the parameters $(H_0, \Omega_{b}\,h^2, \Omega_{dm}\,h^2, A_s, n_s)$\footnote{Although it is more conventional to use $\theta$ instead of $\rm H_0$ when sampling the CMB likelihood, this choice is irrelevant for the present work, as the relative entropy is invariant under parameter transformation (cf. Appendix \ref{sec:entr_props}).}. We also sample the re-ionisation optical depth $\tau$, but marginalise numerically over it before computing the relative entropy, as this parameter is unconstraint by all updates except for Planck 15. To investigate the impact of model extensions, we extend our fiducial model in two ways. We allow either an arbitrary but constant equation of state parameter for Dark Energy (parameterised by $w=P_\text{DE}/\rho_\text{DE}$) or non-flat geometries (parametrised by the fraction of curvature density $\Omega_K = 1-\Omega_\text{tot}$). Furthermore, we also consider the effect of a change of the prior, by using Planck15 constraints as a prior. In the following, we present and interpret the results.

\subsection{Fiducial Configuration: WMAP9 Prior in flat $\Lambda$CDM}

\begin{figure}[t]
    \centering
    \includegraphics[width=\textwidth]{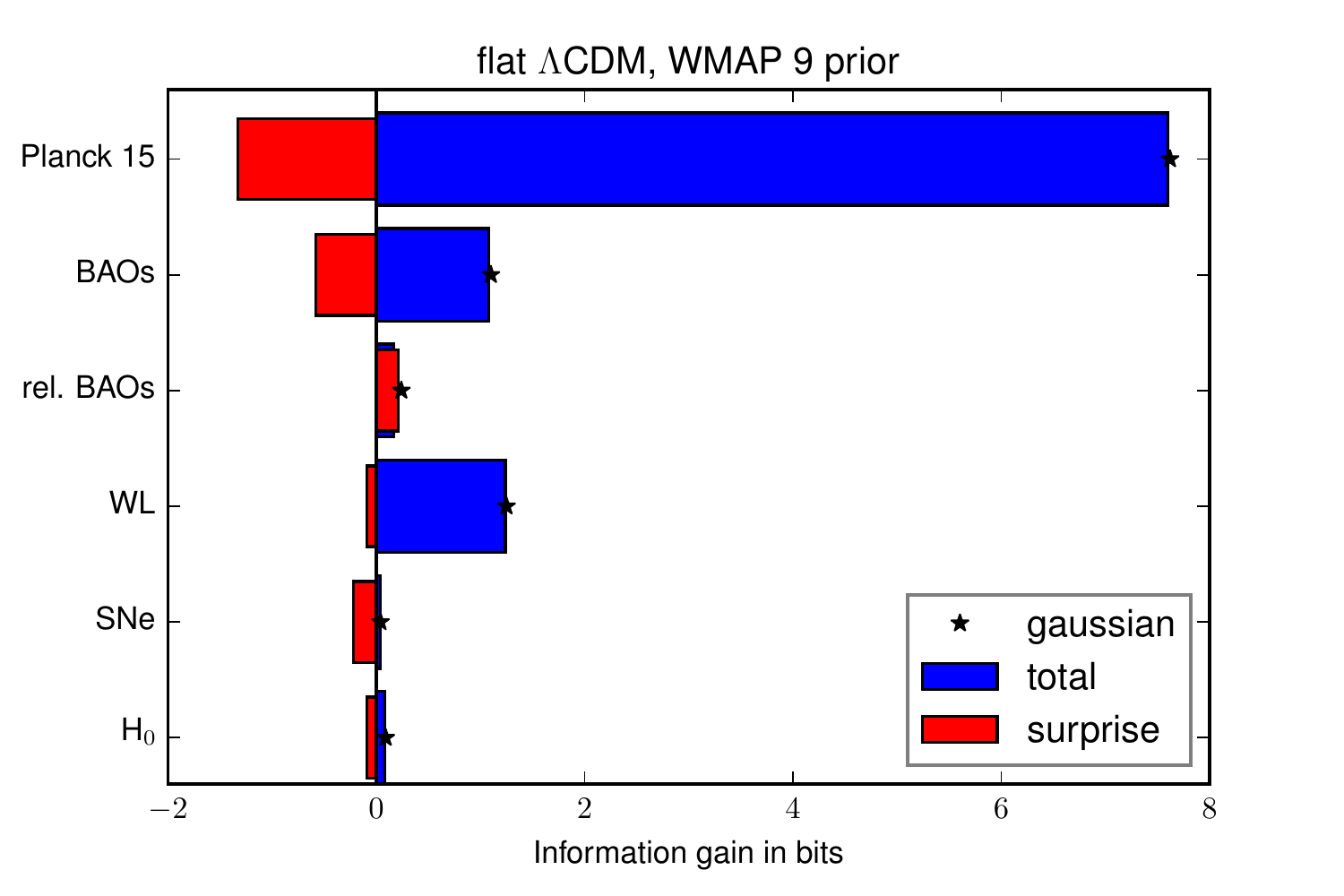}
    \caption{\label{fig:info_gain_wmap} Information gained when updating WMAP9 with different cosmological probes in flat $\Lambda$CDM. The blue bar indicates the information gain computed with the general estimator, the black star the analytic expression for the gaussian case. The red bar shows the surprise.}
\end{figure}

We first consider the information gains in our fiducial configuration: WMAP9 priors within a flat $\Lambda$CDM cosmology. Figure~\ref{fig:info_gain_wmap} shows these gains, with the corresponding numerical values given in Table~\ref{sec:full_wmap}. We see that the dominant additional information comes from the Planck15 data with 7.6 bits (shown by the blue horizontal bar). In red, we show the contribution to the relative entropy coming from the surprise. To calculate the surprise we assume that the prior and the posterior are multivariate Gaussians, thereby allowing us to decompose the information gain into the components shown in equation~\ref{eq:gauss_entr}. In the case of Planck15, we obtain $S = -1.3$ bits of negative surprise. A negative surprise indicates that the mean values of the posterior deviate from the prior by less than expected. From Table~\ref{sec:full_wmap}, however, we see that this level of surprise is not significant, as the expected variation in the relative entropy from statistical fluctuations in the data, $\sigma(D) = 4.7$, is larger than the observed surprise. The results presented here differ from the values presented by \citeR{seehars2}, as we marginalised over $\tau$. In light of the large positive surprise value found by \citeR{seehars1} between WMAP 9 and Planck 13, \citeR{seehars2} discussed the statistical problem of measuring concordance on the example of CMB surveys.

The stars in Figure~\ref{fig:info_gain_wmap} show the Gaussian approximation calculation of the information gain and can be compared to our full numerical results that are shown by the blue bar. For the present case of flat $\Lambda$CDM with WMAP9 prior, the prior and the posteriors are well described by the multivariate Gaussian approximation, leading to good agreement with our numerical calculations. 

After Planck15, we find that the BAO and weak lensing experiments generate the greatest information gains with 1.1 and 1.2 bits, respectively. Comparing the information gain from BAOs with relative BAOs we find that there is a significant drop in information, down to 0.2 bits, when only using relative BAO measures. Since the difference between these two calculations is whether the BAO scale is predicted by the cosmological model or allowed to vary independently, the strength of BAOs rests on the fact that measured scales can be related to the BAO scale from the CMB era through the cosmological model. The final two probes that we consider, $H_0$ and SNe, bring little extra information (0.04 and 0.08 bits respectively) in the flat $\Lambda$CDM case. A final interesting feature that is worth noting from the results in Figure~\ref{fig:info_gain_wmap} is that in most cases we see a small negative surprise. None of these surprises are statistically significant, but a tendency to be negative means that the updated results tend to agree with the prior to a higher degree than expected.

\subsection{Model Extensions}

Within a given model, the relative entropy, and hence the information gain, is invariant under reparkametrizations. However, the relative entropy depends on the model, the prior, the likelihood and hence the data. In model extensions, a parameter which was fixed in the fiducial model is allowed to vary in the extended model. In extended models, the constraints often show degeneracies between the new and some old parameters, for example the geometrical degeneracies between $H_0$ and $\Omega_K$ for the CMB (see e.g.~\citeR{howlett}). If the update breaks such a degeneracy (e.g. by constraining either $H_0$ or $\Omega_K$), its information gain in the extended model will be larger than in the fiducial model. This comes from the fact that the probe carries information that helps to constrain the new free parameter. In the extended model, this information gain is expressed by the tightening of contours in parameter space. In the fiducial model, however, this information gain is suppressed by the theory prior which fixes the parameter. 

On the other hand, if the update does not constrain the new parameter, the relative entropy in both models will remain the same as the new parameter can be marginalised over, leaving the entropy unchanged (see Section~\ref{sec:theor_back} and Appendix~\ref{sec:entr_props}). As a consequence, the relative entropy does not increase. This means, that it cannot decrease under extension of the model: it either increases or is unchanged, as we indeed find for the models and the updates we considered (cf. Tables~\ref{sec:full_wmap}, \ref{sec:full_omk}, and \ref{sec:full_w}). In the first case, the update carries information constraining the new parameter. In the second case, we say that the information gain \emph{saturated}, i.e. that the information carried by the update is already expressed in the unextended model so that the theory prior does not erase information contained in the update. 

In principle, the total information carried by a probe could be measured by dropping all theory priors on parameters of the model and just considering the changes in the space of observables, as argued by~\citeR{modelbreaking}. In practice, however, we expect the information gain to saturate after some few successive model extensions, as wider and wider degeneracies open. Thus, among the information gains computed in different parameter spaces for the same probe, the largest one is the best approximation to the total information gain. By reporting this value, we assess the contribution of each probe to our knowledge of the cosmological parameters. Naturally, this assessment is limited to the parameter spaces we explored and to the data sets we used.

\subsubsection{WMAP9 Prior in non-flat $\Lambda$CDM}

\begin{figure}[t]
    \centering
    \includegraphics[width=\textwidth]{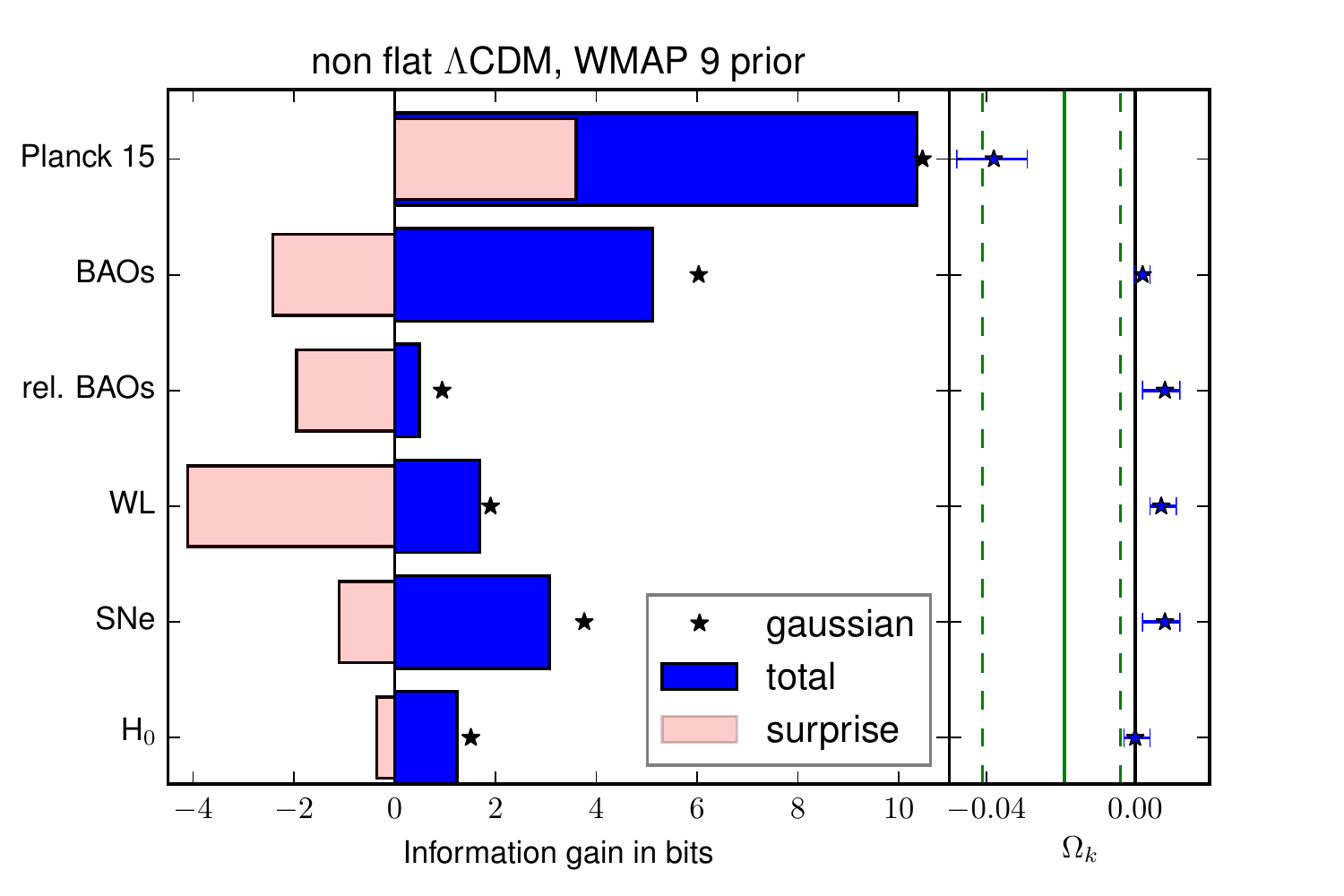}
    \caption{\label{fig:info_gain_wmap_Ok} \emph{Left panel:} Information gained when updating WMAP9 with different cosmological probes in a non-flat $\Lambda$CDM cosmology. The blue bar indicates the total information gain computed with our numerical estimator, the star the value in the Gaussian approximation. The faded red bar shows the surprise. \emph{Right panel:} Medians and 1-$\sigma$ percentiles of the constraints on $\Omega_K$: the green line indicate the WMAP9 median, the dashed green line indicate its 1-$\sigma$ percentiles. The solid black line indicates flatness. All probes are consistent with WMAP9 and, besides Planck15, all probes are consistent with a flat geometry.}
\end{figure}

Figure~\ref{fig:info_gain_wmap_Ok} shows the relative entropy results for the case where we move away from our fiducial model and allow for curvature. Interestingly, we see that, for this more flexible model, the information generated by the individual probes is different than for our fiducial case. In particular we see SNe, H$_0$ and BAO measures now generate 3.1, 1.2 and 5.1 bits of information, respectively, which is significantly larger than for the flat case. By contrast, Planck15\footnote{Regarding Table~\ref{sec:full_omk} we see that in non flat $\Lambda$CDM about $3.6$ bits of information gain of Planck15 are due to surprise. Thus, the information gain due to tightening of contours is $6.4$ bits, and therefore comparable to the information gain in the fiducial configuration.} and the weak lensing experiments provide the same information as they did in the flat case. This suggests that for the latter two probes, their dominant constraining power is within the flat $\Lambda$CDM parameter subspace. For SNe, we see that the information gain of $3.1$ bits when curvature is allowed is a large improvement compared to the 0.08 bits in the flat case. This means that SNe data provides information in the parameter direction which is opened up by allowing for non-flat geometries. The theory prior when moving to flat models (a $\delta$-function in the $\Omega_K$ direction) thus dominates completely over probes that are able to improve curvature constraints such as SNe and {$\rm H_0$} measurements.

For the non-flat $\Lambda$CDM model we again find that the dominant update to WMAP9 comes from the addition of Planck15. The rank order of the other probes becomes BAOs, SNe, $H_0$, weak lensing and relative BAOs. In the case of the BAOs, we see once more that it is not only the late time measurement of a fixed scale that matters, but being able to link this scale through theoretical calculation to the cosmological parameters. This is shown by the large gain of 5.1 bits (up from 1.1 bits for flat $\Lambda$CDM) for the BAOs, while the gain from relative BAOs stays small at 0.5 bits.

In Figure~\ref{fig:info_gain_wmap_Ok}, we also show the Gaussian approximation calculations (star and faded red horizontal bar). We find that, for this extended model, the WMAP9 prior and some of the posteriors are not very well described by multivariate Gaussian distributions. Thus, the surprise results should be seen as approximate and should not be over interpreted. We nevertheless find that the surprise tends to be negative for most probes, indicating stronger agreement with the prior than expected. Planck15 shows positive surprise ($S = 3.6$ bits), but even if the Gaussian approximation were to hold, this would not be statistically significant.

On the right panel of Figure~\ref{fig:info_gain_wmap_Ok} we also show the one dimensional errors on curvature after marginalising over the other parameters. We find that, with the exception of Planck15, all results are consistent with the $\Omega_K = 0$ flat model at $95\%$ confidence. As also found by the Planck collaboration~\cite[pag. 38]{planck15}, we see that the Planck15 data alone favours a closed model.

\subsubsection{WMAP9 Prior in flat $w$CDM}\label{sec:info_gain_wCDM}

\begin{figure}[t]
    \centering
    \includegraphics[width=\textwidth]{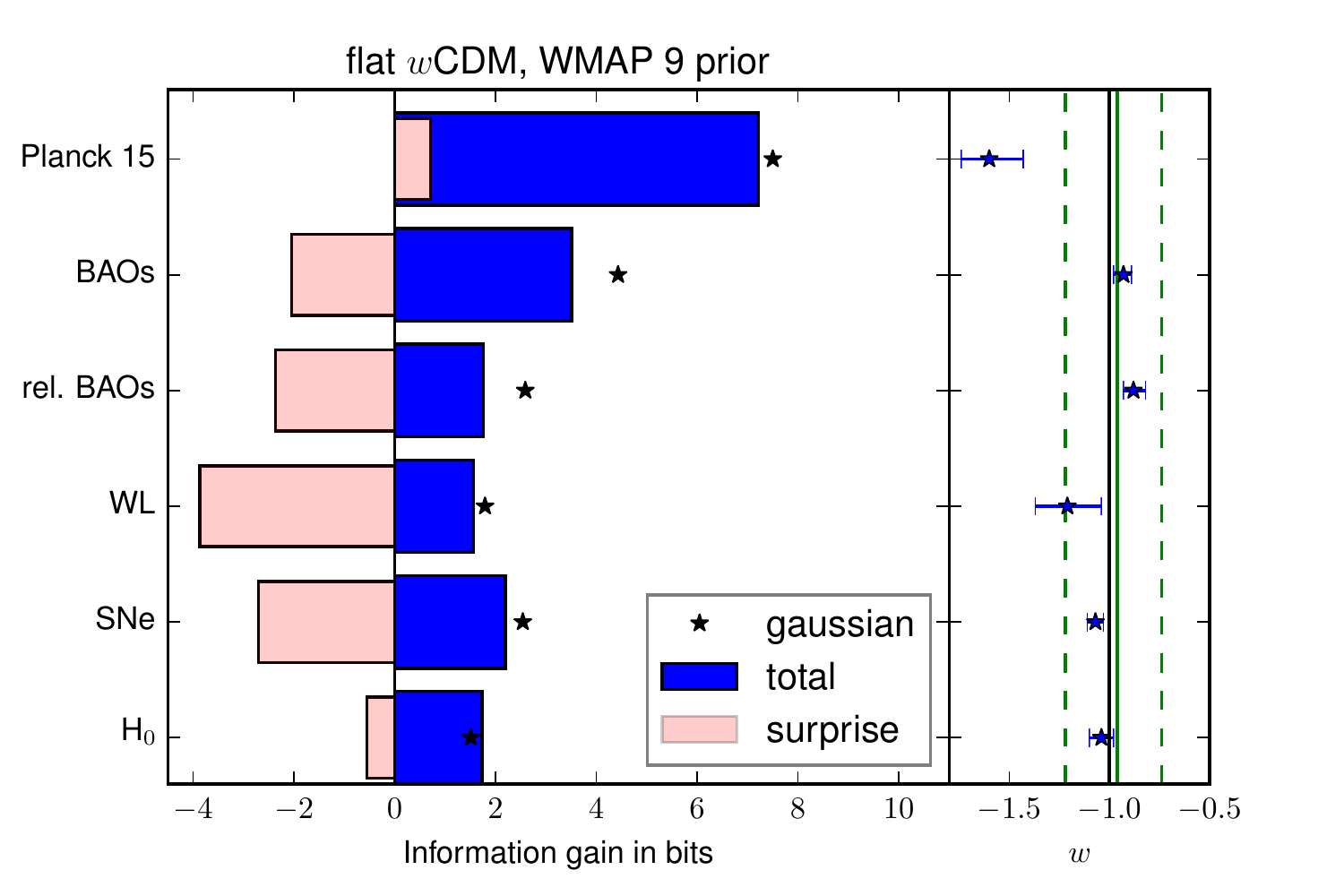}
    \caption{\label{fig:info_gain_wmap_w} \emph{Left panel:} Information gain when updating WMAP9 with different cosmological probes in the flat $w$CDM case. The blue bar indicates the total information gain computed with our numerical estimator, the black star the value in the Gaussian approximation. The faded bar shows the surprise term. \emph{Right panel:} Medians and 1$\sigma$ percentiles of the $w$ constraints. The solid green line indicates the WMAP9 constraints, the dashed, green line the 1$\sigma$ percentiles. The solid black line indicates $w=-1$, i.e. a cosmological constant. All probes except Planck15 are consistent both with WMAP9 and with a cosmological constant. Planck15 prefers Phantom Dark Energy, $w<-1$.}
\end{figure}

The second extension beyond our fiducial model consists in allowing for an arbitrary but constant equation of state parameter $w$ for the Dark Energy component. Figure~\ref{fig:info_gain_wmap_w} shows the information gains for flat $w$CDM. We find that the contributions from the low-redshift probes are more similar to each other (and even to the CMB) than for the other models. These findings are perhaps not surprising since these probes have been, for the most part, developed to constrain the Dark Energy sector. We see again that the inclusion of Planck15 data has the biggest impact. All probes show negative surprise, but once again caution is needed since the Gaussian approximation used for the calculation of this term does not hold well for this extended model, as the CMB prior deviates significantly from a multivariate Gaussian distribution. This manifests in the offset between the black stars and the blue bars in Figure~\ref{fig:info_gain_wmap_w}. A new feature that is more prominent here compared to the other models is the fact that relative BAO measures now deliver a significant amount of information; more than weak lensing and $\rm H_0$ and comparable to SNe experiments. In this case, the fact that we are able to measure a fixed scale at different redshifts in the late time Universe, where Dark Energy dominates, is sufficient to make substantial constraints even without the link between this scale and the cosmological parameters or any knowledge of its intrinsic size.

In the right panel of Figure~\ref{fig:info_gain_wmap_w}, we show the one-dimensional marginal errors on the constant equation of state parameter $w$ when each of the probes is combined with WMAP9. Similar to our earlier results, we find that all probes except Planck15 are consistent with an equation of state of $w = -1$, our fiducial $\Lambda$CDM model, at $95\%$ confidence. We see that Planck15 alone tends to favour a phantom Dark Energy equation of state ($w < -1$), as discussed by Ref.~\cite[pag. 39]{planck15}. 

\subsection{Planck15 Prior in flat $\Lambda$CDM}\label{sec:res_planck}

\begin{figure}[t]
    \centering
    \includegraphics[width=\textwidth]{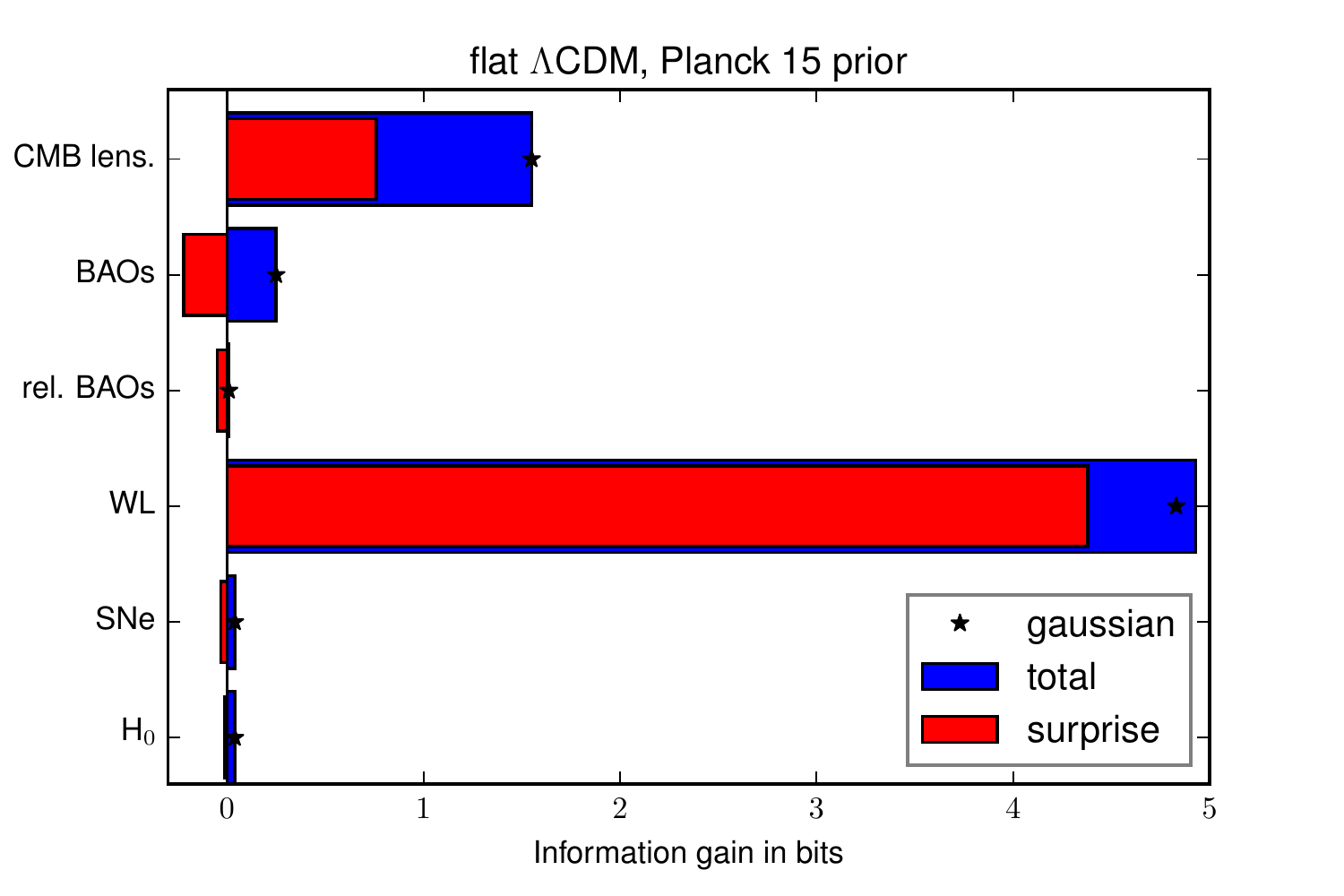}
    \caption{\label{fig:info_gain_planck} Information gained when updating Planck15 with different cosmological probes in flat $\Lambda$CDM. The blue bar indicates the information gain computed with the general estimator, the black star the analytic expression for the gaussian case. The red bar shows the surprise.}
\end{figure}

As a final excursion from our fiducial configuration, we explore the information gained when the Planck15 constraints on flat $\Lambda$CDM are used as the prior. For this purpose, we use the public chains provided by the Planck team~\citeR{planck15}. Figure~\ref{fig:info_gain_planck} shows our results. As was the case when using the WMAP9 prior in the flat $\Lambda$CDM, we see that SNe, ${\rm H_0}$ and relative BAO measure provide little additional information. We also find that the additional gain from the full BAO constraints is reduced from 1.1 bits when WMAP9 is used as a prior to 0.25 bits when using Planck15. All the geometrical probes also show small levels of negative surprise. 

The most striking feature of Figure~\ref{fig:info_gain_planck} is the information gain when CFHTLenS weak lensing results are combined with Planck15. We see that the results are strongly dominated by a positive surprise. This shows that the weak lensing update has moved the means more than statistically expected. This level of surprise, 4.4 bits, is an $8\sigma$ effect and is thus very significant. Our findings are in line with \citeR{2015arXiv151000688R}, which finds a "substantial" disagreement between the temperature and polarisation measurements of Planck 15 and the CFHTLenS constraints using an evidence ratio based method and ultra conservative cuts for the CFHTLenS data. 

The significance of $8\sigma$ might at first seem to disargee with other work \cite{2015arXiv151000688R, maccrann, WL_systematics, WL_systematics_3}, which finds tensions of the order of $3\sigma$, as can be already guessed by visual inspection of the marginalised constraints in the $\Omega_M$, $\sigma_8$ plane. The $3\sigma$ reported in the literature refer to distances of mean values in units of standard deviations, which is another quantity than the significance of a surprise, as discussed in Section \ref{sec:gauss_entr}. Furthermore, as reported already by \citeR{2015arXiv151000688R}, and discussed at length in \citeR{seehars2}, inspection of marginalised contours can lead to an underestimation of the degree of tension between data sets. One should therefore be cautious to infer the significance of a tension from the distance in means of the marginalised contours. An example for the subtle effects that can be caused by correlations in high dimensional constraints is the comparison between WMAP and Planck 2013 constraints in \citeR{seehars2}. The correlated constraints on the cosmological parameters between those two posteriors show no significant tension in any of the one-dimensional marginal distributions. Yet, when reparametrizing such that the WMAP constraints become uncorrelated, a $5 \sigma$ tension in one of the new parameters appears. The surprise is invariant under such transformations and hence detects this deviation already in the untransformed cosmological parameters. 

The surprise we find when updating Planck15 by CFHTLenS data could indicate the presence of residual systematics in the data or it could be a consequence of new physics. \citeR{maccrann} investigated these possibilities and showed that the tension between Planck CMB and CFHTLenS is affected by various weak lensing systematics, but could not be resolved by considering neutrino masses, tensor modes, or massive sterile neutrinos. However, \citeR{dark_current} claims that the tension between CFHTLenS and Planck 15 temperature data can be resolved by assuming a new form of dark current interacting with dark matter. Recent works by \citeR{WL_systematics, WL_systematics_3} have shown that the agreement between CFHTLenS and Planck is sensitive to the treatments of the systematics effects in the WL data and non linearities in the small angular scales, and can be explained almost exclusively by more careful analysis of the CFHTLenS data. 

The significant tension between Planck15 and our simple implementation of CFHTLenS measurements suggests that further investigations are needed. Recent weak lensing results from the Dark Energy Survey~\cite{2015arXiv150705552T}, for example, seem to achieve better agreement. In our opinion, it would also be interesting to apply the surprise measure to the revised CFHTLenS constraints \cite{WL_systematics, WL_systematics_3}. This would give a statistically robust, multidimensional assessment of the degree of tension present between the revisited data set and the Planck15 CMB measurements. Once again the calculation of surprise depends on the approximation that prior and posterior are Gaussian, which in this case we find to agree very well.

Figure~\ref{fig:info_gain_planck} also shows the impact of CMB lensing. In this work, we used the constrains computed by~\citeR{planck_lensing}, where a CMB lensing reconstruction based on the Planck15 temperature and polarisation anisotropy measurements~\cite{planck15} was performed. We see that this additional probe gives an improvement (1.6 bits) that is greater than the other probes, not including the weak lensing results. However, as with our earlier results this may not continue to hold if we were to consider extended models. Concerning the CMB lensing result, we find that half of the information gain comes from the surprise term, 0.8 bits. However, being only a 1-$\sigma$ deviation, this surprise is not statistically significant (cf. Table \ref{sec:full_planck}).

\section{Conclusion}\label{sec:conc}

In this paper, we compare constraints on three cosmological models (flat $\Lambda$CDM, non-flat $\Lambda$CDM, and flat $w$CDM) from $H_0$, SNe, BAO, weak lensing, and CMB data. For this purpose we extend work by \citeR{seehars1}, where the information gains on flat $\Lambda$CDM constraints from a historical sequence of CMB data were quantified by the Kullback-Leibler divergence, or relative entropy, of the individual posterior distributions. This measure allows us to quantify how much information is contributed to the constraints of the parameters by individual probes and how this information content depends on the model. The relative entropy also provides a way of quantifying the tension introduced by a Bayesian update. It measures tension in the same unit as the information yielded from the improvement in the precision of the constraints. Thus, our proposed measure of tension, the surprise, can be directly compared to the information gain by a cosmological probe, as already shown by \citeR{seehars1, seehars2}. 

It is interesting to note that the measured information gain of a given probe depends on the theoretical priors imposed on the model. For instance, the information gain of a probe that is sensitive to curvature will be reduced if we consider a flat model. More generally, imposing theory priors will tend to suppress the measured information gain. As an approximation to the total information contained in each probe, we hence consider the maximal information gain in the extended models covered in this work.

In order to rank the experiments by their contributions to cosmology relative to WMAP9, we therefore calculate their maximal information gain over all considered models. Using this definition, we find that the following experiments have the greatest contribution to cosmology: Planck (10 bits), followed by BAO surveys (5.1 bits) and SNe experiments (3.1 bits). We also find that other cosmological probes, including weak lensing (1.7 bits) and {$\rm H_0$} measures (1.7 bits), bring information but at a lower level.

Considering WMAP9 priors, we do not find any significant surprises for any of the probes in any of the models. This means that the individual experiments are in good agreement with WMAP9. Furthermore, considering model extensions, we find that all the low redshift probes are consistent with a flat $\Lambda$CDM model. Our results highlight the fact that the strengths of these probes are not to be found in their constraining power on flat $\Lambda$CDM parameters, but in their potential to detect deviations from this model in extended scenarios.

Planck15 constraints do not display any significant surprise either when compared to WMAP9, even though we also considered the Planck 15 polarisation data, as also reported by \citeR{seehars2}. However, as was also found by the Planck collaboration~\cite{planck15}, we see that the 2015 Planck release of CMB observations tends to push the constraints away from a flat $\Lambda$CDM cosmology, centering instead on models with curvature and phantom Dark Energy. Considering that the addition of complementary data, among which CMB lensing measurement of Planck itself, shifts the central values of the Planck constraints by $2\sigma$, as reported by \citeR{planck15}, it might be interesting to compute the surprise values for the different updates to Planck 15. This would allow us to spot any tension between the CMB and low redshift observations in extended models. 

Furthermore, we quantify the information gained by adding the low redshift probes to a Planck15 prior in flat $\Lambda$CDM. We find that adding weak lensing data from CFHTLenS offers substantial information (4.8 bits) but that most of this is dominated by a surprise (4.4 bits with a significance of $8\sigma$). This robustly quantifies the known tensions between Planck and CFHTLenS~\cite{maccrann, 2015arXiv151000688R} which have become alleviated by recent weak lensing measurements from the Dark Energy Survey~\cite{2015arXiv150705552T} and recent revisions of the CFHTLenS analysis by \citeR{WL_revised, WL_systematics, WL_systematics_3}.

Our results highlight the versatility of the relative entropy as a figure of merit for the constraining power of cosmological probes and for detecting tensions between them. Our information gain estimates confirm the CMB as the dominant source of cosmological information. However, as we allow for cosmologies beyond flat $\Lambda$CDM, the other probes already achieve information gains comparable to those by the CMB. With many future large-scale structure surveys on the way it will be interesting to see how the landscape of cosmological probes changes in the next decade.

\acknowledgments

This work was in part supported by the Swiss National Science Foundation (Grant No. $200021\_143906$). We thank Joel Akeret for his support with CosmoHammer. We thank the anonymous referee for the useful comments. We also, thank Adam Riess for the input on the present status of distance ladder measurements. We acknowledge use of the Planck Legacy Archive. Planck ($\texttt{http://www.esa.int/Planck}$) is an ESA science mission with instruments and contributions directly funded by ESA Member States, NASA, and Canada. This work is partially based on observations obtained with MegaPrime/MegaCam, a joint project of CFHT and CEA/IRFU, at the Canada-France-Hawaii Telescope (CFHT) which is operated by the National Research Council (NRC) of Canada, the Institut National des Sciences de l'Univers of the Centre National de la Recherche Scientifique (CNRS) of France, and the University of Hawaii. This research used the facilities of the Canadian Astronomy Data Centre operated by the National Research Council of Canada with the support of the Canadian Space Agency. CFHTLenS data processing was made possible thanks to significant computing support from the NSERC Research Tools and Instruments grant program. We acknowledge the use of the Legacy Archive for Microwave Background Data Analysis (LAMBDA), part of the High Energy Astrophysics Science Archive Center (HEASARC). HEASARC/LAMBDA is a service of the Astrophysics Science Division at the NASA Goddard Space Flight Center. 

\bibliographystyle{JHEP}
\bibliography{my_bib}

\appendix
\section{Shapelet Reconstruction}\label{sec:shapelets}
Following the reasoning of \citeR{shapelets}, we project a given PDF $f(\pmb{x})$ on the orthonormal, complete basis (ONB) given by the eigenstates $\phi_{\pmb{m}}(\pmb{x})$ of the $d$ dimensional Quantum Harmonic Oscillator. These functions are called shapelets. As they are multidimensional Hermite polynomials times a Gaussian, expanding in shapelet basis amounts to computing the Gram-Charlier Series \cite{gram, charlier1, charlier2, hermites} given by

\begin{equation}\label{eq:gram_charlier}
f(\pmb{x}) = \sum_{\pmb{m} \in  \mathbb{N}^d} f_{\pmb{m}} \,\phi_{\pmb{m}}(\pmb{x}). 
\end{equation}
As the shapelets form an ONB, we can easily compute 

\begin{equation}
f_{\pmb{m}} = \int_{\mathbb{R}^d} f(\pmb{x})\, \phi_{\pmb{m}}(\pmb{x})\, d^d\pmb{x} = \langle \phi_{\pmb{m}}(\pmb{x}) \rangle_f.
\end{equation}
The $\pmb{m}$th shapelet coefficient is the expected value of the $\pmb{m}$th shapelet on the distribution $f$. The expected value $\langle \phi_{\pmb{m}}(\pmb{x}) \rangle_f$ and its statistical error $\sigma(f_{\pmb{m}})$ can be estimated with a Monte Carlo integral from any sample of $f$. This requires an evaluation algorithm for $\phi_{\pmb{m}}(\pmb{x})$ based on recurrence relations of one dimensional shapelets (e.g. \citeR{shapelets}). Thereafter, the signal to noise of detection $\frac{\rm S}{\rm N} (f_{\pmb{m}})$ of each shapelet coefficient can be computed. Applying a signal to noise cut $\nu$, allows us to terminate the series \eqref{eq:gram_charlier}, and thus, to compress the PDF into some few shapelet coefficients.

As for galaxy images, faster convergence in shapelet coefficients can be obtained be choosing a suitable shapelet center and shape. As center we utilise the mean $\pmb{\mu}_f$ estimated from the sample. For the shape of the shapelets, we take account of the covariance $C_f$ by defining elliptical shapelets as

\begin{equation}
B^f_{\pmb{m}} (\pmb{x}):= \frac{1}{|C_f|^{1/4}} \, \phi_{\pmb{m}} \left\{ C_f^{-1/2}(\pmb{x}-\pmb{\mu}_f) \right\}, 
\end{equation}
where $|C_f|$ denotes the determinant of $C_f$. The additional factor is needed to ensure the orthonormality of the elliptical shapelets. Consequently, the algorithm described above can be also used to estimate the coefficients in elliptical shapelet basis.

Besides the signal to noise cut $\nu$, we introduce another tuning parameter: we scale the covariance matrix $C_f$ by a factor $\lambda$, imitating the shapelet scale used by \citeR{shapelets} for galaxy images. Adapting the shapelet scale improves the reconstruction. To asses the optimal choice of tuning parameters $\nu$ and $\lambda$, we try to maximise the likelihood that the sample $\pmb{x}_i$ has been drawn from $\hat f$, the reconstruction of $f$, given by

\begin{equation} \label{eq:selfentropy}
\ln P(\pmb{x}_i | \hat f) = \ln  \prod_i \hat f(\pmb{x}_i)= \sum_{i=1}^N \ln  \hat f(\pmb{x}_i) \approx N \big \langle \ln \hat f \big \rangle_f.
\end{equation}
This gives as best fit values for the tuning parameters $\nu$ and $\lambda$. For the CMB degeneracies we encountered in the extended models we find  $\nu_\text{best fit} \approx 10$ and $\lambda_\text{best fit} \approx 1.1$. Thus, by maximising \ref{eq:selfentropy}, an optimal shapelet reconstruction can be computed.

This yields a set of coefficients $f_{\pmb{m}}$, which given \eqref{eq:gram_charlier} can be used as to estimate $f$ on any nodes $\pmb{x}$. However, any finite series of shapelets will have some negative values. Nevertheless, these negative points lie far away from the sample, where any kernel density estimator is inherently inaccurate. As these values are close to zero, we just output their absolute value to ensure positiveness.

\section{Properties of the Relative Entropy} \label{sec:entr_props}

\paragraph{Almost Positiveness} $D(p||q) \geq 0$ for all  $p(\pmb{x})$ and $q(\pmb{x})$ and  $D(p||q) = 0$ if and only if $p(\pmb{x})= q(\pmb{x})$ almost everywhere (for a proof see~\citeR{kullbackleibler}).

\paragraph{Invariance under reparametrisation} Take any bijective, measurable mapping $\pmb{\psi}$ of the parameters, s.t. $\pmb{y}=\pmb{\psi}(\pmb{x})$. It maps $p(\pmb{x}) \mapsto p^\prime (\pmb{y})= \det \big( \frac{d\pmb{\psi}}{d\pmb{x}}\big) \, p(\pmb{\psi}({\pmb{x})})$ and consequently $p(\pmb{x})\, d^d\pmb{x} = p^\prime(\pmb{y})\, d^d\pmb{y}$. Given the transformation above, it follows from \eqref{eq:def_rel_ent} that 

\begin{equation}
D(p(\pmb{x})||q(\pmb{x})) = D(p^\prime (\pmb{y})||q^\prime (\pmb{y})),
\end{equation}
i.e. the relative entropy is \emph{invariant under transformations} in parameter space. This property is especially useful when comparing different cosmological probes, e.g. it is irrelevant if the information gain of CFHTLenS is computed with $A_S$ or with $\sigma_8$ as power spectrum normalisation.

\paragraph{Compatibility with marginalisation} Assume that prior and posterior differ only in one parameter, as happens, for instance, when a CMB experiment is updated by a local $H_0$ measurement. Then $p(H_0, \pmb{x}) = l(H_0)\, q(H_0, \pmb{x})$, where $\pmb{x}$ are some other parameters which are not constrained by the update. For the relative entropy this means 

\begin{equation}
    \begin{split}
         D(p||q) &= \int dH_0\, d^d \pmb{x} \, p(H_0, \pmb{x}) \ln(l(H_0)) = \\
         &= \int dH_0 \, p_{\text{marg}}(H_0) \ln(l(H_0)) =\\
         &= D(p_{\text{marg}}(H_0)||q_{\text{marg}}(H_0)),
    \end{split}
\end{equation}
where $p_{\text{marg}}(H_0) = \int d^d \pmb{x}\, p(H_0,\pmb{x})$ is the marginalised distribution. Hence, the relative entropy is invariant under marginalisation of unconstrained parameters. When updating WMAP9 by an H$_0$ measurement, for example, the information gain can be computed solely in the marginal distribution of the parameter $H_0$. This significantly reduces the dimensionality of the computation.

\section{Full Table of Information Gains}\label{sec:full}
In this Appendix we provide the full tables of all information gains. Following the derivation of~\citeR{seehars1}, we present the entropy in the Gaussian approximation $D$, the expected entropy $\langle D \rangle$, the surprise $S = D - \langle D \rangle$, and the standard deviation of the entropy $\sigma(D)$ derived for the Gaussian case. Furthermore, we present the numerical values $\hat D$ computed with the scheme presented in Section~\ref{sec:gener_entr}. $\dag$ indicates the use of publicly available chains for the posterior. $\ddag$ indicates the use of the `replacing data scheme'; otherwise the `additional data scheme' is used (for details see~\citeR{seehars1}).

\subsection{Flat $\Lambda$CDM, WMAP9 as Prior}\label{sec:full_wmap}

Parameters: $(H_0, \Omega_{b}\,h^2, \Omega_{dm}\,h^2, A_s, n_s)$

\vspace{10pt}
\begin{tabular}{|cccccc|}
  \hline
  Probe & $D$ & $\langle D \rangle$ & $S$ & $\sigma(D)$ & $\hat D$\\
  \hline
  SNe & 0.09 & 0.18 & -0.09 & 0.22 & 0.08 $\pm$ 0.01\\
  H0 & 0.04 & 0.26 & -0.22 & 0.31 & 0.04 $\pm$ 0.01\\
  BAOs & 1.10 & 1.68 & -0.58 & 0.92 & 1.08 $\pm$ 0.01\\
  rel. BAO's & 0.24 & 0.03 & 0.21 & 0.04 & 0.17 $\pm$ 0.01\\ 
  WL & 1.25 & 1.34 & -0.09 & 0.88 & 1.24 $\pm$ 0.02\\
  Planck15 \dag \ddag & 7.62 & 8.59 & -1.33 & 4.71 & 7.60 $\pm$ 0.03\\ 
  \hline
\end{tabular}

\subsection{Non-flat $\Lambda$CDM, WMAP9 as Prior}\label{sec:full_omk}

Parameters: $(H_0, \Omega_{b}\,h^2, \Omega_{dm}\,h^2, A_s, n_s, \Omega_K)$

\vspace{10pt}
\begin{tabular}{|cccccc|}
  \hline
  Probe & $D$ & $\langle D \rangle$ & $S$ & $\sigma(D)$ & $\hat D$\\ \hline
  SNe & 3.76 & 4.86 & -1.10 & 7.23 & 3.07 $\pm$ 0.07\\
  H0 & 1.21 & 1.57 & -0.36 & 1.15 & 1.23 $\pm$ 0.01\\
  BAOs & 6.03 & 8.43 & -2.42 & 3.24 & 5.12 $\pm$ 0.01\\
  rel. BAOs & 0.94 & 2.89 & -1.95 & 3.14 & 0.49 $\pm$ 0.01\\
  WL & 1.90 & 6.01 & -4.11 & 4.83 & 1.69 $\pm$ 0.02\\
  Planck15 \dag \ddag & 10.47 & 6.87 & 3.60 & 2.43 & 10.36 $\pm$ 0.03\\ 
  \hline
\end{tabular}

\subsection{Flat $w$CDM, WMAP9 as Prior}\label{sec:full_w}

Parameters: $(H_0, \Omega_{b}\,h^2, \Omega_{dm}\,h^2, A_s, n_s, w_0)$

\vspace{10pt}
\begin{tabular}{|cccccc|}
  \hline
  Probe & $D$ & $\langle D \rangle$ & $S$ & $\sigma(D)$ & $\hat D$\\ \hline
  SNe  & 2.54 & 5.45 & -2.70 & 3.31 & 2.19 $\pm$ 0.01\\
  H0  &  1.51 & 2.06 & -0.55 & 1.08 & 1.73 $\pm$ 0.01\\
  BAO & 4.43 & 6.48 & -2.05 & 3.19 & 3.51 $\pm$ 0.02\\
  rel. BAO's & 2.59 & 4.94 & -2.37 & 3.66 & 1.76 $\pm$ 0.88\\
  WL & 1.79 & 5.66 & -3.87 & 4.53 & 1.56 $\pm$ 0.02\\
  Planck15 \dag \ddag & 7.22 & 6.51 & 0.71 & 1.90 & 7.50 $\pm$ 0.03\\
  \hline
\end{tabular}

\subsection{Flat $\Lambda$CDM, Planck15 as Prior}\label{sec:full_planck}

Parameters: $(H_0, \Omega_{b}\,h^2, \Omega_{dm}\,h^2, A_s, n_s)$

\vspace{10pt}
\begin{tabular}{|cccccc|}
  \hline
  Probe & $D$ & $\langle D \rangle$ & $S$ & $\sigma(D)$ & $\hat D$\\ \hline
  SNe & 0.04 & 0.05 & -0.01 & 0.06 & 0.04 $\pm$ 0.01\\
  H0  & 0.04 & 0.07 & -0.03 & 0.09 & 0.04 $\pm$ 0.01\\
  BAOs \dag& 0.25 & 0.47 & -0.22 & 0.12 & 0.25 $\pm$ 0.02\\
  rel. BAO's & 0.01 & 0.06 & -0.05 & 0.01 & 0.01 $\pm$ 0.01\\ 
  WL & 4.83  & 0.45 & 4.38 & 0.50 & 4.93 $\pm$ 0.02\\
  CMB lensing \dag & 1.55 & 0.79 & 0.76 & 0.67 & 1.55 $\pm$ 0.01\\
  \hline
\end{tabular}

\section{Impact of different $\rm H_0$ measurements}\label{sec:h0}
Different values for the Hubble constant have been proposed after the recalibration of the geometrical distance to NGC 4258 by \citeR{humphrey}. \citeR{efstathiou} reported $H_0^\text{Efstathiou} = 70.6 \pm 3.3 \text{ km/s/Mpc}$, and \citeR{riess14} measured $H_0^\text{Riess} = 73.0 \pm 2.4 \text{ km/s/Mpc}$. The smaller error of $H_0^\text{Riess}$ comes from the fact, that it considers two further distance anchors \cite{riessmail}.

\begin{table}[h] \label{tab:h0}
\centering
\caption{Results for two different $\rm H_0$ values in the different models we considered.}
\vspace{10pt}
\begin{tabular}{|cccccc|}
\hline
flat $\Lambda$CDM, WMAP9 priors & $D$ & $\langle D \rangle$ & $S$ & $\sigma(D)$ & $\hat D$\\ \hline
\hline
$H_0^\text{Efstathiou}$ & 0.04 & 0.26 & -0.22 & 0.31 & 0.04 $\pm$ 0.01\\ 

$H_0^\text{Riess}$ & 0.37 & 0.38 & -0.01 & 0.42 & 0.34 $\pm$ 0.02 \\ \hline
\hline

non-flat $\Lambda$CDM, WMAP9 priors & $D$ & $\langle D \rangle$ & $S$ & $\sigma(D)$ & $\hat D$\\ \hline
\hline
$H_0^\text{Efstathiou}$ & 1.21 & 1.57 & -0.36 & 1.15 & 1.23 $\pm$ 0.01\\ 

$H_0^\text{Riess}$ & 1.22 & 1.31 & -0.08 & 1.22 & 1.84 $\pm$ 0.01 \\ \hline
\hline

flat $w$CDM, WMAP9 priors & $D$ & $\langle D \rangle$ & $S$ & $\sigma(D)$ & $\hat D$\\ \hline
\hline
$H_0^\text{Efstathiou}$ & 1.51 & 2.06 & -0.55 & 1.08 & 1.73 $\pm$ 0.02\\ 

$H_0^\text{Riess}$ & 1.85 & 2.49 & -0.64 & 1.05 & 2.10 $\pm$ 0.01 \\ \hline
\hline

flat $\Lambda$CDM, Planck 15 priors & $D$ & $\langle D \rangle$ & $S$ & $\sigma(D)$ & $\hat D$\\ \hline
\hline
$H_0^\text{Efstathiou}$ & 0.04 & 0.07 & -0.03 & 0.09 & 0.04 $\pm$ 0.01\\ 

$H_0^\text{Riess}$ & 0.20 & 0.08 & 0.12 & 0.11 & 0.20 $\pm$ 0.01 \\ \hline

\end{tabular}
\end{table}

Table \ref{tab:h0} summarises the results obtained by using the two different values for $\rm H_0$. We conclude that $H_0^\text{Riess}$ gives a larger information gain (2.1 bits) than to $H_0^\text{Efstathiou}$ (1.7 bits). This is not only due to the higher precision, but also to the less negative surprises. However, this difference does not change much of the main conclusions of this work, exemplifying how the specific choice of the calibration of a dataset does not change the main ranking of probes. Anyways, it is worth noting that we find a positive surprise between $H_0^\text{Riess}$ and Planck 15, as already noted by \citeR{planck15}. This surprise is just slightly larger than $1-\sigma$ and therefore not significant.

\section{Impact of the Planck TE and EE spectra}\label{sec:planck_pol}

We check how much this result may depend on the fact that we also considered the Planck 15 polarisation data. To this purpose we compute the information gained from adding the TE and EE constraints, representing the small scale polarisation data, to the TT\_lowTEB, consisting of the large scale polarisation data and the temperature measurements. We found 0.7 bits in flat $\Lambda$CDM, 1.4 bits in non-flat $\Lambda$CDM and 0.8 bits in flat $w$CDM. As for H$_0$, this has no significant impact on the overall ranking, but would just affect the specific numerical values.

The addition of polarisation data results in negative surprises at a significance value just above 1-$\sigma$: -1.1 bits in flat $\Lambda$CDM ($\sigma (D)=1.0$), -1.0 bits in non-flat $\Lambda$CDM ($\sigma (D)=1.2$) and -1.2 bits in flat $w$CDM ($\sigma (D)=0.9$). This means that adding the Planck 15 polarisation data to the Planck 15 temperature data shifts the mean less than expected. Therefore, we do not expect any significant changes in our surprise results when omitting the polarisation data.

\end{document}